\documentclass{ws-procs975x65}

\usepackage{bbm}
\DeclareFontFamily{OT1}{rsfs}{}
\DeclareFontShape{OT1}{rsfs}{m}{n}{ <-7> rsfs5 <7-10> rsfs7 <10->rsfs10}{}
\DeclareMathAlphabet{\mycal}{OT1}{rsfs}{m}{n}
\newcommand{\lan}{{\mycal L}}

\newcommand{\vp}{\mathbf{p}}

\newcommand{\dual}[1]{\overset{{}^{{}^{\boldsymbol{\neg}}}}{\smash[t]{#1}}}

\begin{document}

\title{Dark Spinors}

\author{C. B. BOEHMER$^*$ and J. BURNETT$^{**}$}

\address{Deparment of Mathematics and Institute of Origins,\\ University College London,\\
London, GU21 6NS, UK\\
$^*$E-mail: c.boehmer@ucl.ac.uk\\
$^{**}$E-mail: j.burnett@ucl.ac.uk}

%

\begin{abstract}
This article will provide the reader a short introduction to dark spinors, which 
are ELKO spinors, eingenspinors of the charge conjugation operator, applied to dark matter/energy. 
\end{abstract}

\keywords{Spinors; cosmology; dark matter; dark energy; elko;}

\bodymatter

\section{INTRODUCTION}
\label{aba:intro}

ELKO spinors are eigenspinors of the charge conjugation operator~\cite{jcap}, they belong to a wider class of so-called flagpole spinors~\cite{daRocha:2005ti}. According to the Wigner classification they are non-standard spinors and obey the unusual property $(CPT)^2=-\mathbbm{1}$. Hence, their dominant coupling to other fields is via the Higgs mechanism or via gravity~\cite{jcap}. The particles associated with such a field theory are dark and therefore it is natural to apply them to the dark matter and dark energy problem, calling them dark spinors henceforth.

Dark spinors are defined by
\begin{align}
      \lambda = \begin{pmatrix} \pm \sigma_2 \phi^{\ast}_{L} \\
                \phi_L \end{pmatrix} \,,
      \label{eq:i1}
\end{align}
where $\phi^{\ast}_{L}$ denotes the complex conjugate of $\phi_L$ and $\sigma_2$ denotes the second Pauli matrix. For a detailed treatment of the field theory of the eigenspinors of the charge conjugation operator we refer the reader to Refs.~\citen{jcap}. Dark spinors have an imaginary bi-orthogonal norm with respect to the standard Dirac dual $\bar{\psi} = \psi^{\dagger}\gamma^0$, and in order for a consistent field theory to emerge the dual is given by
\begin{align}
       \dual{\lambda}_u = i\,\varepsilon_u^v \lambda_v^{\dagger} \gamma^0 \,,
       \label{eq:i3}
\end{align}
with $\varepsilon_{\{+,-\}}^{\{-,+\}}=-1=-\varepsilon_{\{-,+\}}^{\{+,-\}}$ such that
\begin{align}
      \dual{\lambda}_u(\vp) \lambda_v(\vp) = \pm\, 2m\, \delta_{uv} \,,
      \label{eq:i4}
\end{align}
where $\vp$ denotes the momentum.

Due to their formal structure, dark spinors couple differently to gravitation than scalar fields or Dirac spinors~\cite{Boehmer:2006qq}, eigenspinors of the parity operator. As stated before this allows for many interesting applications, for instance, in Ref.~\citen{Boehmer:2007ut} it has been shown that dark spinors naturally yield an anisotropic expansion in the context of cosmological Bianchi type I models. This allows for a suppression of the low multipole amplitude of the primordial power spectrum. The primordial power spectrum of the quantum fluctuations of dark spinors has been investigated in Refs.~\citen{Boehmer:2008rz,Gredat:2008qf} where it was found that the small scale power spectrum essentially agrees with that of scalar field inflation while the large scale power spectrum shows new features. Causal propagation issues have recently been discussed in Refs.~\citen{Fabbri:2009ka}.

We have also shown that dark spinors could be considered a candidate particle to describe dark energy~\cite{Boehmer:2009aw}. Their equation of state, $w$, is dynamical and exhibits phantom properties (= crosses $w=-1$) with late time convergence to cosmological constant $w=-1$. Finally, detailed proofs of the results shown in this proceeding can be found in Ref.~\citen{Boehmer:2008ah}.

\section{THE GOVERNING EQUATIONS}

The following construction is done in the Einstein-Cartan setting, this is the same as saying our full connection $\tilde{\Gamma}$ is made up of the Levi-Civita connection $\Gamma$ and a contorsion part $K$, $ \tilde{\Gamma}^\lambda_{\mu\nu} = \Gamma^\lambda_{\mu\nu} - K_{\mu\nu}{}^{\lambda}$. The action of Einstein-Cartan gravity is
\begin{align}
      S = \int \Bigl( \frac{M_{\rm pl}^2}{2} \tilde{R} + \tilde{\lan}_{\rm mat} \Bigr)
      \sqrt{-g}\, d^4 x,
      \label{eq:a1}
\end{align}
where $\tilde{R}$ is the Ricci scalar computed from the complete connection with contortion contributions, $g$ is the determinant of the metric, $\tilde{\lan}_{mat}$ denotes the matter Lagrangian and $1/M_{\rm pl}^2 = 8\pi G$ is the coupling constant; the speed of light is set to one $(c=1)$. The resulting field equations are
\begin{align}
      \tilde{G}_{ij} = \tilde{R}_{ij} - \frac{1}{2} \tilde{R} g_{ij} &= \frac{1}{M_{\rm pl}^2}\,\Sigma_{ij},
      \label{eq:t8} \\
      T^{ij}{}_{k} + \delta^i_k T^{j}{}_{l}{}^{l} - \delta^j_k T^{i}{}_{l}{}^{l}
      &= M_{\rm pl}^2\,\tau^{ij}{}_{k},
      \label{eq:t9}
\end{align}
where $\tau^{ij}{}_{k}$ is the spin angular momentum tensor, defined by
$ \tau_{k}{}^{ji} = \frac{\delta \tilde{\lan}_{\rm mat}}{\delta K_{ij}{}^{k}},$
and $\Sigma_{ij}$ is the total energy-momentum tensor
$
      \Sigma_{ij} = \sigma_{ij} + (\tilde{\nabla}_k-K_{lk}{}^{l})
      (\tau_{ij}{}^{k}-\tau_{j}{}^{k}{}_{i}+\tau^{k}{}_{ij}),
      \label{eq:t10}
$
where $\sigma_{ij}$ is metric energy-momentum tensor
\begin{align}
  \sigma_{ij} = \frac{2}{\sqrt{-g}} \frac{\delta(\sqrt{-g}\tilde{\lan}_{\rm mat})}{\delta g^{ij}}. 
\end{align}
The field equations~(\ref{eq:t9}) are in general 24 algebraic equations, and in the absence of spin sources torsion vanishes, torsion does not propagate.


\section{RESULTS}

One finds that the solution to these equations for the dark spinor is,

\begin{align}
      \frac{\dot{\varphi}}{\varphi} = -\frac{\sqrt{V_0/M_{\rm pl}^2}}{4\sqrt{3}}
      \frac{8 + 3 \varphi^4/M_{\rm pl}^4}{12 -\varphi^4/M_{\rm pl}^4}
      \sqrt{4 - \varphi^4/M_{\rm pl}^4}.
      \label{eq:dy2}
\end{align}

\begin{figure}[!htb]
\centering
\epsfig{file=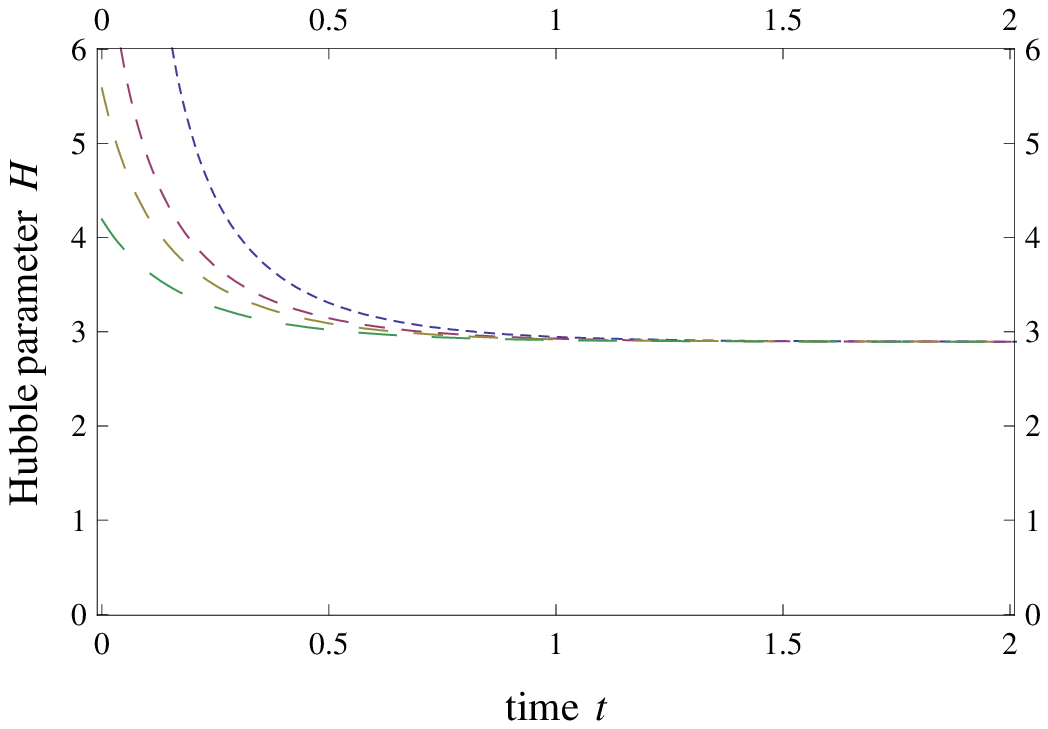,width=2in}
\epsfig{file=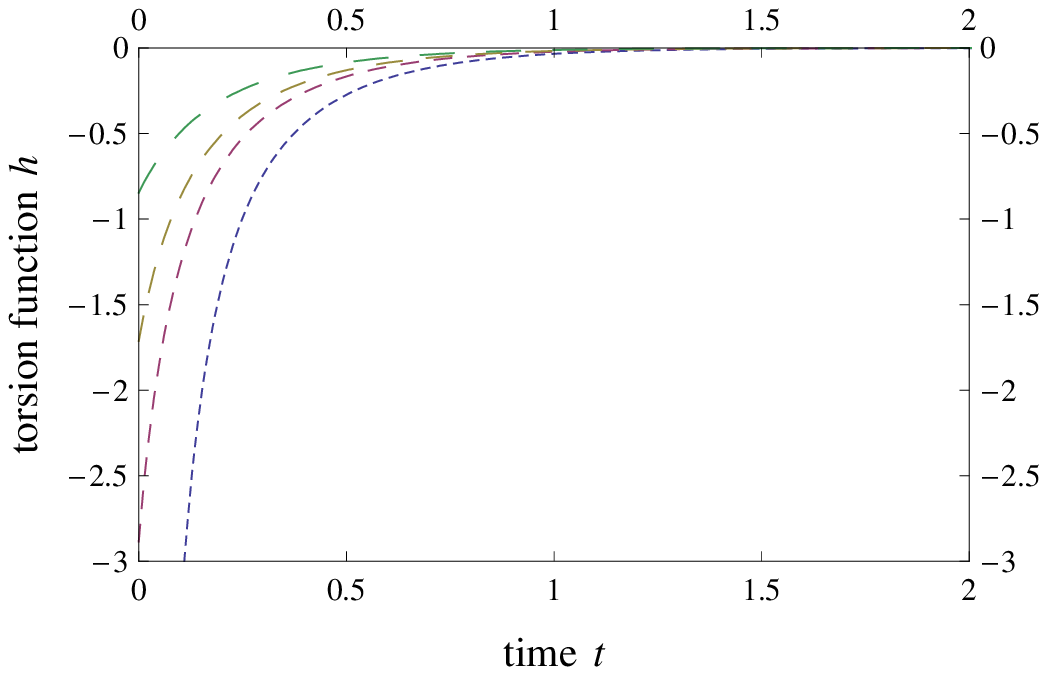,width=2in}

\caption{Left: Hubble parameter and right: torsion function $h$ for $1/M_{\rm pl}^2 = 8\pi$ and $V_0 = 1$. Initial conditions of the matter field are $\varphi_i=\varphi(t=0)=\{0.282,0.25,0.23,0.20\}$, $\{$blue (short dashed), red (dashed), (medium dashed) yellow and green (long dashed) $\}$}.
\label{fig1}
\end{figure}

Once one has (\ref{eq:dy2}) it is possible to investigate how both the Hubble parameter and torsion evolve as they both depend on the dark spinor. With some simple calculations one can obtain the graphs in Fig.~\ref{fig1}.

\section{DISCUSSION}

It is important to realise that the above construction cannot be achieved with Dirac spinors, the spin angular momentum trivially goes to zero in a cosmological setting. Therefore if torsion is discovered to be present in the universe then Dirac spinors will not suffice. This paper shows, at least as a source of torsion, that dark (ELKO) spinors are a good alternative. Moreover due to their construction and the heavy constraint on their interaction with the electromagnetic force, they are naturally dark, and therefore could be considered for dark matter. It was also mentioned and we refer the reader to Ref.~\citen{Boehmer:2009aw}, that one can demonstrate in a normal Einstein universe (= curvature only), that dark spinors exhibit properties of dark energy. Their {\it dynamical} equation of state converges to $w=-1$ and also permits crossing into the phantom regime ($w<-1$) with convergence still holding. When a dark spinor crosses into the phantom regime it does not create ghosts and therefore does not suffer from the same problem as previous phantom models. It is also possible for one to show that a dynamical dark energy model based on dark spinors requires their potential to be of the simplest form, namely a canonical mass term, without self-interaction.

\bibliographystyle{ws-procs975x65}

\end{document}